# A Statistical Test for Clades in Phylogenies

Thurston H. Y. Dang[1], and Elchanan Mossel[2]

[1]*Department of Electrical Engineering and Computer Sciences, University of California, Berkeley, CA, 94704, United States of America*

[2]*Department of Statistics, The Wharton School, University of Pennsylvania*

*Pennsylvania PA, 19104, United States of America*

**Corresponding author:** Thurston Dang, Department of Electrical Engineering and Computer Sciences, 387 Soda Hall, Berkeley, CA, 94704, United States of America; E-mail: thurstond@berkeley.edu

*Abstract.* We investigated testing the likelihood of a phylogenetic tree by comparison to its subtree pruning and regrafting (SPR) neighbors, with or without re-optimizing branch lengths. This is inspired by aspects of Bayesian significance tests, and the use of SPRs for heuristically finding maximum likelihood trees.

Through a number of simulations with the Jukes-Cantor model on various topologies, it is observed that the SPR tests are informative, and reasonably fast compared to searching for the maximum likelihood tree. This suggests that the SPR tests would be a useful addition to the suite of existing statistical tests, for identifying potential inaccuracies of inferred topologies.

(Keywords: clades, phylogenies, Bayesian, SPR, pruning, regrafting)

There are variety of likelihood-based tests for phylogenies (see Goldman, Anderson, & Rodrigo, 2000 for a review). These are useful because, when the topology is questionable, there are often "several reasonable trees" with similar likelihoods (Yang, Goldman, & Friday, 1995).

Furthermore, these tests are practical, because it is possible to obtain such comparison trees through minor modifications to the tree of interest (Foster, 2001), and calculating the likelihood of an arbitrary tree (given the data) can be efficiently performed with dynamic programming (Felsenstein, 1981). For example, FastTree applies the Shimodaira-Hasegawa test on the three alternate topologies obtained by nearest-neighbor interchanges (NNIs) (Price, n.d.), while PhyML (Guindon, 2010) performs an approximate likelihood ratio test (aLRT), comparing the inferred topology with the second-best topology obtained by an NNI.

## *A statistical test for clades*

We begin with an intuitive derivation of our test. Let:

- $T$ be the "true" tree
- $D$ be the data generated from the tree
- $\hat{T}$ be the tree obtained by maximum likelihood from $D$ (Fig. 1)
- $S$ be the set of all tree topologies that have the same number of species as $T$
- $R \subseteq S$
- $g$ be the branch lengths
- $G(g)$ be a prior over the branch lengths
- $g' = \underset{g}{\mathrm{argmax}}\, P(D|\hat{T}, g)$

The probability that a tree is correct given the sequence data can be calculated with Bayes' rule (e.g., Velasco, 2008):

$$P(T = \hat{T}|D) = \frac{\int_G P(D|\hat{T}, g)P(\hat{T})}{\sum_{t \in S} \int_G P(D|t, g)P(t)}$$

This can be simplified by assuming a uniform prior over the trees (a very common assumption; see Velasco, 2008):

$$P(T = \hat{T}|D) = \frac{\int_G P(D|\hat{T}, g)}{\sum_{t \in S} \int_G P(D|t, g)}$$

One approach for calculating the posterior probability is to use Markov Chain Monte Carlo (MCMC), integrating over nuisance parameters such as branch lengths (e.g., Huelsenbeck & Imennov, 2002). We instead restrict our comparison trees to a subset $R \subseteq S$:

$$P(T = \hat{T}|D) \leq \frac{\int_G P(D|\hat{T}, g)}{\sum_{t \in R} \int_G P(D|t, g)}$$

and if we assume we have an abundance of data, the integrated likelihood is similar to the maximum:

$$\approx \frac{G(g')P(D|\hat{T}, g')}{\sum_{t \in R} G(g')P(D|t, g')}$$

$$= \frac{P(D|\hat{T}, g')}{\sum_{t \in R} P(D|t, g')} \tag{1}$$

if we are using the same branch lengths $g'$ for all trees.

The branch lengths $g'$ are not optimal for every topology. Hence, we can get a better bound by re-optimizing the branch lengths for every tree in the denominator i.e.,

$$P(T = \hat{T}|D) \leq \frac{G(g')P(D|\hat{T}, g')}{\sum_{t \in R} \max_g G(g)P(D|t, g)}$$

$$= \frac{P(D|\hat{T}, g')}{\sum_{t \in R} \max_g P(D|t, g)} \quad (2)$$

if we assume a uniform prior over the branch lengths.

*Subtree pruning and regrafting.* – The extraordinary number of possible topologies $((2n - 5)!!$; e.g., Felsenstein, 1978), which made it impractical to exhaustively search for the maximum likelihood topology, necessitates that our set of comparison topologies, $R$, must also be drastically smaller. To serve as the "several reasonable trees" (Yang, Goldman, & Friday, 1995), we propose using the topologies obtained by subtree pruning and regrafting (SPRs). For example, in Figure 2a, one possible subtree to prune is {Z,E,F}, which can be regrafted onto any of the six remaining edges, such as DY, WX or AW (Fig. 2b). When the substitution model is reversible, the tree is unrooted (e.g., Felsenstein, 1981), which means we could also prune the subtree consisting of all nodes to the left of the midpoint YZ, onto the edges EZ or FZ (Fig. 2c).

Using SPRs has a number of advantages:

- The resulting statistical test has an intuitive interpretation: the alternative topologies are obtained by moving a "clan" (a clade/subtree under some possible rooting; e.g., Zhu, Degnan, & Steel, 2011) to another location. For example, in Figure 3, where the subtree

"below" Z is attached to BC with a long branch, the likelihood may be similar or even improved when regrafted to another edge such as AB, CD, DE or EF.

Matching this intuition, many phylogenetic analysis programs rely on a series of SPRs to "intensively search the tree space" (Hordijk & Gascuel, 2005).

- SPRs model recombination or horizontal gene transfer (e.g., Allen & Steel, 2001).
- SPRs are tractable: they are quadratic in the number of species (see Table 1). This is far fewer than the number of possible topologies (Table 2).

Let $SPR(X)$ be {all new topologies reachable from $X$ by exactly one SPR operation} ∪ {$X$}. Accordingly, our statistics corresponding to equations (1) and (2) are:

- $SPR_{plain}$ statistic: $\dfrac{P(D|\hat{T},g\prime)}{\sum_{t \in SPR(\hat{T})} P(D|t,g\prime)}$

- $SPR_{opt}$ statistic: $\dfrac{P(D|\hat{T},g\prime)}{\sum_{t \in SPR(\hat{T})} \max_g P(D|t,g)}$

Both statistics are only intended to provide an approximate upper bound on the likelihood that the topology is correct.

For the $SPR_{plain}$ statistic, since the branch lengths are not re-optimized for the trees in the denominator, there are four copies of each Type (ii) topology (when ignoring branch lengths), with possibly different branch lengths (Fig. 4). Since we have no prior information, we weight each of those topologies by ¼. Similarly, since we do not know where along the destination edge to regraft the tree, we use the midpoint of the destination edge.

Although $SPR_{opt}$ gives a better bound than $SPR_{plain}$, and some authors have cautioned against not re-optimizing parameters (Goldman, Anderson, & Rodrigo, 2000), it may be less desirable due to the extra run-time incurred (although partially offset by only needing ¼ of the Type (ii) topologies). We therefore evaluate both SPR$_{opt}$ and SPR$_{plain}$ for accuracy and speed.

Our statistics are perhaps most similar to Aris-Brosou's (2003) Bayes significance test (BST):

$$\log BF_{\bar{T},i} \approx \frac{1}{k}\sum_{j=1}^{k} \log p(X|T_j) - \log p(X|T_i)$$

However, the BST uses some different approximations (e.g., a geometric average in place of an arithmetic average, which allowed for the summation of log-likelihoods), is intended to provide decisive evidence if the Bayes factor ($BF_{\bar{T},i}$) is sufficiently large, and does not specify the nature of the comparison trees $T_j$.

*Limitations.* – Suppose the likelihood of all trees is shown as per Figure 5, and that $SPR(R) = \{M, ..., W\}$, $SPR(F) = \{A, ..., J\}$. (For concreteness, we refer to SPR neighborhoods, but the arguments apply to any symmetric neighborhood.)

Ideally, the likelihood for R would be calculated as $\frac{\Pr(D|Tree_R)}{\sum_{i\in\{A,...,W\}} \Pr(D|Tree_i)} = 0.348$. Our approximation using only the SPR neighborhood yields the bound $\frac{\Pr(D|Tree_R)}{\sum_{i\in\{M,...,W\}} \Pr(D|Tree_i)} = 0.455$.

Similarly, the true likelihood for $F$ would be 0.139; using only the SPR neighbors gives 0.595, higher than for the maximum likelihood tree. While this does not violate the property that our SPR statistics provide only an upper bound on the likelihood of correctness, it is undesirable.

For this example, $F$ is the only pathological case, since every other tree would have $F$ and/or $R$ as an SPR neighbor (by symmetry), and therefore have low likelihood according to the test. However, we would expect larger topologies to have many light-tailed local optima: Guindon (2010) had previously observed that the aLRT (which is based on NNIs) performed poorly if a good topology had not been found.

## METHOD

### *Implementation*

We implemented the SPR tests with Perl and shell scripts. We used Seq-Gen (Rambaut & Grass, 1997) to generate data sets from the trees, and DNAML (Felsenstein, 2005) to heuristically search for the maximum likelihood tree.

### *Scenarios*

For all scenarios, we assumed the Jukes-Cantor model for both data generation and phylogenetic inference. With the Jukes-Cantor model, the equilibrium base probabilities $\pi_A = \pi_C = \pi_G = \pi_T$, and $P_{ij}(v) = \begin{cases} \frac{1}{4} + \frac{3}{4} e^{\frac{-4v}{3}} & \text{if } i = j \\ \frac{1}{4} - \frac{1}{4} e^{\frac{-4v}{3}} & \text{if } i \neq j \end{cases}$ (e.g., Felsenstein, 2008).

*Accuracy of the SPR tests.* – Define $B_i$ to be the fully balanced binary tree with $2^i$ species. Let $C_i$ be $B_i$ but with an additional two species sub-tree, attached to the root via a branch of length two. All branch lengths not otherwise specified are 0.05. $C_1$, $C_2$, and $C_3$ are shown in Figure 6.

We generated data from the following trees:

- 4-species tree (Fig. 4a), with length 2 for branch XY, and unit length for all other branches: 100,000 simulations

- $C_2$ (6-species): 10,000 simulations
- $C_2$ but where the longer branch is 2.5: 10,000 simulations
- $C_4$ (18-species): 1000 simulations

The numbers of simulations stated above were used for calculating the average likelihood of obtaining the correct tree by ML, and the $SPR_{plain}$ statistics. Since the $SPR_{opt}$ statistics are slower to calculate, they are generally averaged over 1000 simulations.

*Speed of the SPR tests.* – To provide a rough estimate of the speed of the phylogenetic analyses and SPR tests, we ran them on $C_3$, $C_4$, $C_5$, and $C_6$ for a single simulation. This also provides some anecdotal evidence of the behavior of the SPR test for larger trees.

*Neighborhoods.* – We noted in the Introduction that using the SPR neighborhood, rather than all topologies, as the denominator could result in some non-ML topologies receiving a high test score. This phenomenon is less likely to occur in small trees because the SPR neighborhood is a substantial subset of all topologies (with 4 species, the SPR neighborhood is all other topologies; with 5 species, the SPR neighborhood contains 12 of the 14 other topologies); however, it is computationally infeasible to investigate large trees. Thus, we investigated the 6-species topology $C_2$, for which the SPR neighborhood contains 30 of the 104 other topologies.

If the sequences are extremely short, then most topologies will be of roughly equal likelihood and no topology will have a high test score, while if the sequences are extremely long, then the maximum likelihood topology will be unambiguously selected. We therefore chose a sequence length of 512 (based on the results in Fig. 9), so that there might be multiple highly plausible topologies.

For each of the 105 possible 6-species topologies (denoted $T_{i \in [1..105]}$), we optimized the branch lengths, and ran the SPR tests with that topology in the numerator i.e.,

- $SPR_{plain}$ statistic for $T_i$: $\frac{P(D|T_i, g'_i)}{\sum_{t \in SPR(T_i)} P(D|t, g'_i)}$, where $g'_i = \underset{g}{\mathrm{argmax}}\, P(D|T_i, g)$

- $SPR_{opt}$ statistic for $T_i$: $\frac{P(D|T_i, g'_i)}{\sum_{t \in SPR(T_i)} \max_g P(D|t, g)}$

This simulates the situation where a phylogenetic analysis program chooses one of these topologies, even if it is not actually the maximum likelihood topology. We also recorded the average likelihood for each of the topologies ($P(D|T_i, g'_i)$).

## RESULTS

Note: In all results below, I(T=T') is calculated using DNAML. DNAML does not necessarily (indeed, is unlikely to, for very large trees) find the true maximum likelihood tree.

### *4-species tree with unit length branches*

The likelihood of correctly recovering the correct topology ranges from roughly chance when the sequence length is 1, to almost perfect when there are 1024+ nucleotides (Fig. 7 and 8).

Both the $SPR_{plain}$ and $SPR_{opt}$ statistics are higher for correct than incorrect topologies, and both statistics approach 1 for the correct topology when the sequences are very long, because the likelihood of the alternative (incorrect) topologies decreases.

However, when the $SPR_{plain}$ statistic is applied to incorrect topologies, it is a weaker bound. This is because the branch lengths of the topology in the numerator are optimized, but the branch lengths of the topologies in the denominator are not. We would expect that with longer sequences, the SPR$_{plain}$ statistic would further decrease for incorrect topologies; however, we did not run the simulations because the time required to gather enough data would be excessive (in part due to the longer sequences, but mostly since it becomes extremely rare to find an instance where the maximum likelihood topology is incorrect).

## *C$_2$ (6-species)*

Perhaps the most striking feature of Figure 9 is the non-monotonic likelihood function. In this case, this is likely because of overly short branches in the true tree (Yang, 1995). We share their opinion that "Since there is no hope of recovering the true tree with such short sequences, we can restrict our discussion to relatively large $n$" (p. 693).

The relations between the test scores are similar to the previous 4-species tree: the statistics are higher for the correct topologies compared to the incorrect topologies; and the SPR$_{plain}$ statistics are higher than the corresponding SPR$_{opt}$ statistic.

The SPR$_{opt}$ statistics are similar for both correct and incorrect topologies for sequence lengths < 256, but they begin to diverge for n ≥ 512 (i.e., once the likelihood is well-behaved).

When the long branch is 2.5, the results are also similar, though our test sequences are not long enough to reliably recover the correct topology (Fig. 10).

## *C$_4$ (18-species)*

With this topology, the likelihood of obtaining the correct topology appears much more mundane. The only unusual result is that for sequences of length n ∈ {128, 256, 512}, it appears that the SPR test scores are higher for incorrect than correct topologies (Fig. 11). It may be due to random sampling, though a genuine difference would not be impossible (as illustrated in the "Neighborhoods" section).

## *Larger trees*

Note: the scores and timings in this section are noisy because they are based on a single simulation. Furthermore, computational resources (e.g., CPU, disk and network bandwidth) were shared between multiple users. All timings are based on a single core of an AMD Opteron Processor 8384 (2.7 GHz).

For each topology, the SPR statistics tend to be higher when the sequences are longer (Fig. 12). Additionally, for a fixed sequence length, the SPR scores are higher for the smaller topologies (i.e., larger topologies are harder to resolve).

The run-times for the SPR tests are shown in Figure 13. For any fixed topology, there is a large constant component to the run-time (for example, for $C_6$ with a sequence length of 1, it takes over 20 minutes), but this is an artifact of the current test implementation. Ignoring the shorter sequence lengths, we can see that the asymptotic complexity of the SPR tests is $O(m^3 n)$, where $n$ is the sequence length and $m$ is the number of species. This is expected: there are $O(m^2)$ topologies obtained by SPRs, and evaluating each topology takes $O(mn)$ time. Furthermore, DNAML's branch optimization procedure is $O(mn)$ per tree, as the branches are each optimized separately (Felsenstein, 2008). Overall, the SPR$_{plain}$ run-times compare favorably with the DNAML run-times (Fig. 14), while the SPR$_{opt}$ run-times, although higher, are still reasonable.

## Neighborhoods

Figure 15 plots the average log-likelihood for each of the 105 6-species topologies, given data which are always generated from topology $C_2$. The topologies have been grouped based on nearest-neighbor interchange (NNI) distance from the correct topology: 0 NNIs means the correct topology ($C_2$), 2 NNIs means a Type (iii) SPR away from $C_2$, and 3 NNIs means they are a Type (iii) SPR away from an NNI neighbor of $C_2$. The graph shows that all topologies that are far removed (> 1 SPR) from the correct topology have very low likelihood i.e., the SPR neighborhood (indeed, even the NNI neighborhood) contains most of the likelihood, making it a good approximation of the total likelihood across all topologies.

The average SPR statistics for each of the 105 topologies are shown in Figure 16. This illustrates the weaker bound of the SPR$_{plain}$ statistic: $C_2$ is given a likelihood of 0.6, while four (incorrect) topologies are given a score > 0.3; in total, the SPR$_{plain}$ test assigns a total likelihood of 3.15 across all the topologies. In contrast, the SPR$_{opt}$ statistic has a total likelihood of 1.02.

The statistics show, reassuringly, that none of the 104 incorrect topologies have a higher score than $C_2$; furthermore, the likelihoods of the topologies that are more than an NNI away are negligible.

Note that the test scores for $C_2$ (distance 0, Fig. 16) cannot be directly compared to those in Figure 9. In Figure 9, "SPR$_{plain}$ | T=T'" shows the likelihood score of $C_2$, when $C_2$ was the maximum likelihood tree. In contrast, Figure 16 shows the likelihood score of $C_2$, whereby we always select $C_2$, even if it is not of maximum likelihood.

# DISCUSSION

Our simulations show that the SPR tests – especially $SPR_{opt}$ – are informative, generally providing low scores for incorrect topologies. Importantly, this is true for moderate length sequences, where a statistical test is useful (whereas for extremely short or long sequences, the likelihood of obtaining the correct or incorrect topology respectively is all but assured). Our results also indicate that both SPR tests are tractable, and that the SPR neighborhood is a good source of alternative topologies.

The rigorousness of our experiments has been limited by the large number of possible phylogenetic trees. Firstly, we have only tested specific examples of n-species topologies, with particular branch lengths. Secondly, our "Neighborhoods" test was performed on a 6-species topology (again, only a specific topology): it is likely that our tree is too small to exhibit multiple meaningful maxima.

We were also unable to run multiple simulations for the "Larger trees" studies. However, this is not a concern for potential test users, since they would only need to run the test once per dataset.

It would also be instructive to apply the test on real datasets (for which the true topology is known by other means): Stamatakis (2005) has cautioned that simulated alignment data has an unrealistically strong phylogenetic signal, due to the absence of gaps or sequencing errors. Analyses of real world data would motivate the use of more complicated models of DNA evolution.

There are many other, newer phylogenetic analysis programs than DNAML, though not all of them will readily analyze data with our assumptions (Jukes-Cantor model etc.). Such programs would be better at identifying the "true" maximum likelihood tree, and/or be faster. More importantly, since these programs often efficiently compute SPRs, we can use their techniques to improve our statistical tests. Improving the speed of the test would help not only users, but also make it practical for us to perform more simulations to assess the accuracy of the test.

In Figure 15, we observed that NNIs contained most of the likelihood for the 6-species tree. Unfortunately, using NNIs instead of SPRs has only minor time savings for small trees, and would provide a much poorer bound for larger trees. However, it is possible to obtain most of the benefits of SPRs, with similar cost to NNIs. PhyML switched from NNIs to SPRs, but used a distance-based heuristic to discard poor SPRs, and estimates SPRs locally (Guindon, 2010). Similarly, FastTree2 only looks at $O(n)$ of the $O(n^2)$ best SPRs (Price, Dehal, & Arkin, 2010), and RAxML uses "lazy subtree rearrangements", whereby only the three branches adjacent to the regrafting point are optimized (Stamatakis & Alachiotis, 2010).

Although we have not implemented these SPR optimizations, our results already provide bounds on the speed and accuracy of related tests. For example, using NNIs as the neighborhood without optimizing the branch lengths (ill-advised as this may be), will be faster than $SPR_{plain}$, but less accurate, while $SPR_{plain}$ with lazy subtree rearrangements will have speed and accuracy in between that of $SPR_{plain}$ and $SPR_{opt}$.

Guindon (2010) observed that the aLRT and bootstrap supports should both be used, since they each detect different problems with trees (e.g., aLRT does not consider alternative topologies that are far removed, while the bootstrap can give high scores when there are very short branches, even though these are not supported by substitutions). Similarly, our tests cannot provide a definitive statement of the correctness of a topology, but might often be able to cast doubt on an incorrect topology. Thus, we envision the SPRs not as a replacement, but a complement to existing phylogenetic tree tests; when used in concert with other tests, the SPR tests are an effective and affordable method for testing phylogenies.

## FUNDING

This work was supported by the Office of Naval Research (N000141110140 and N00014-14-1-0823 to E.M.).

## ACKNOWLEDGEMENTS

This paper is based on a master's thesis written by T.D. under the supervision of E.M. at the Department of Statistics at UC Berkeley. T.D. thanks Satish Rao and Allan Sly for helpful comments and suggestions and for serving on his thesis committee.

| Type | Regrafting to an edge ... | # topologies | # unique, new topologies (ignoring branch lengths) |
|---|---|---|---|
| i | adjacent to the cut edge | 6(n-2) | 0 |
| ii | one away from the cut edge | 8(n-3) | 2(n-3) |
| iii | more than one away from the cut edge | 4(n-3)(n-4) | 4(n-3)(n-4) |
| Total | | (2n-3)(2n-4) | 2(n-3)(2n-7) |

**Table 1. From Theorem 2.1 (Allen & Steel, 2001). Type ii sub-tree pruning and regrafting (SPRs) are equivalent to nearest neighbor interchanges (NNIs).**

| # species | # unique topologies | # new, unique SPRs | Type II + III SPRs |
|---|---|---|---|
| 3 | 1 | 0 | 0 |
| 4 | 3 | 2 | 8 |
| 5 | 15 | 12 | 24 |
| 6 | 105 | 30 | 48 |
| 7 | 945 | 56 | 80 |
| 8 | 10,395 | 90 | 120 |
| 9 | 135,135 | 132 | 168 |
| 10 | 2,027,025 | 182 | 224 |
| 18 | $1.92 \times 10^{17}$ | 870 | 960 |
| 34 | $1.12 \times 10^{44}$ | 3782 | 3968 |
| 66 | $1.65 \times 10^{107}$ | 15,750 | 16,128 |

**Table 2. The number of possible topologies and SPRs for n species.**

$T$

$D$

| | |
|---|---|
| A | CGTATGCA |
| B | ACGTCATG |
| C | AAGTGACG |
| E | GTACGAGA |
| F | TTACTGTG |

$\hat{T}$

**Figure 1. The true tree, data generated according to the true tree, and the maximum likelihood tree obtained from the data.**

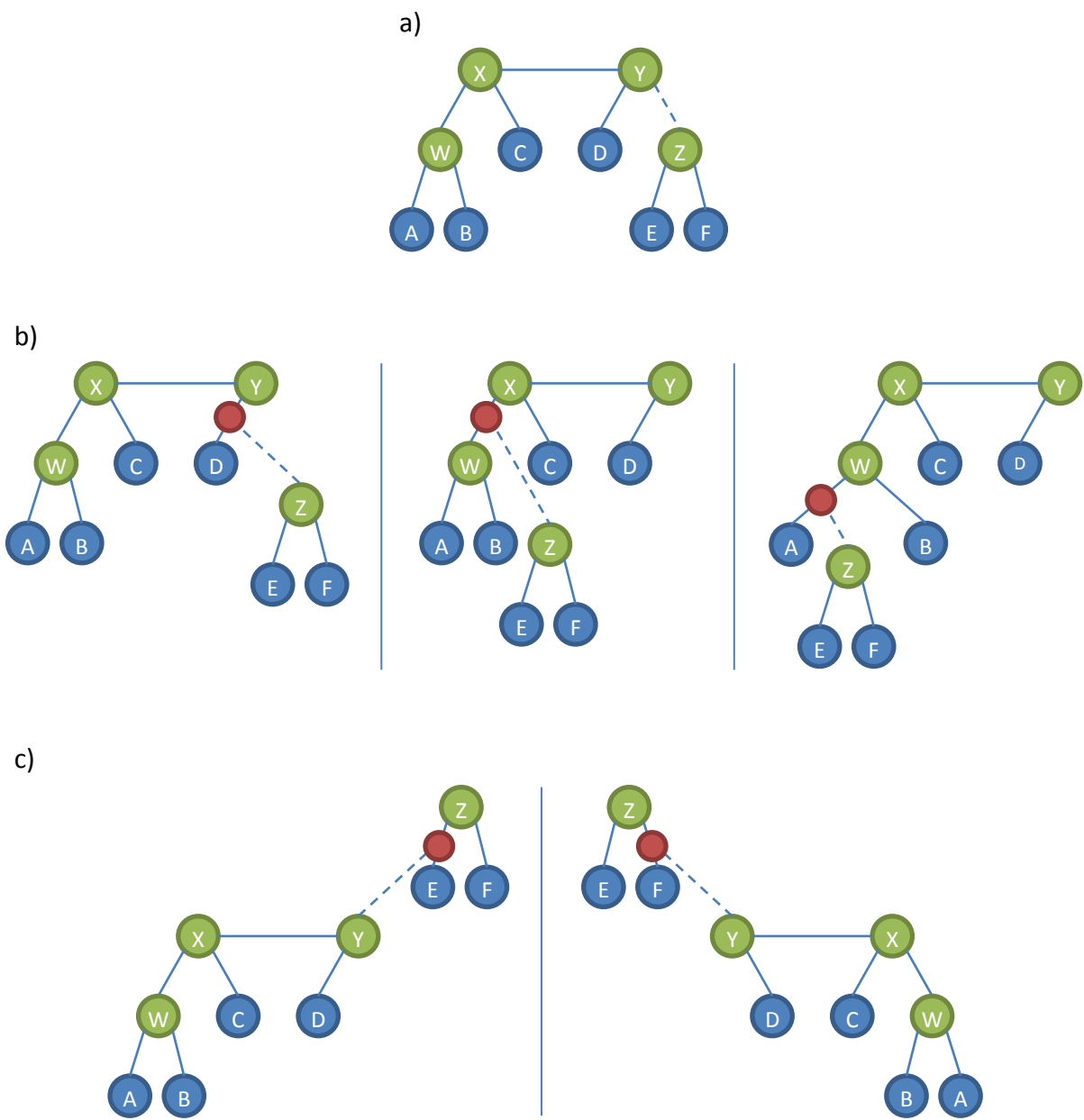

**Figure 2.** a) A topology; b) three (of six) topologies that can each be obtained by pruning the subtree {Z,E,F} and regrafting onto the rest of the tree; c) alternatively, we can regraft the rest of the tree onto the subtree {Z,E,F}.

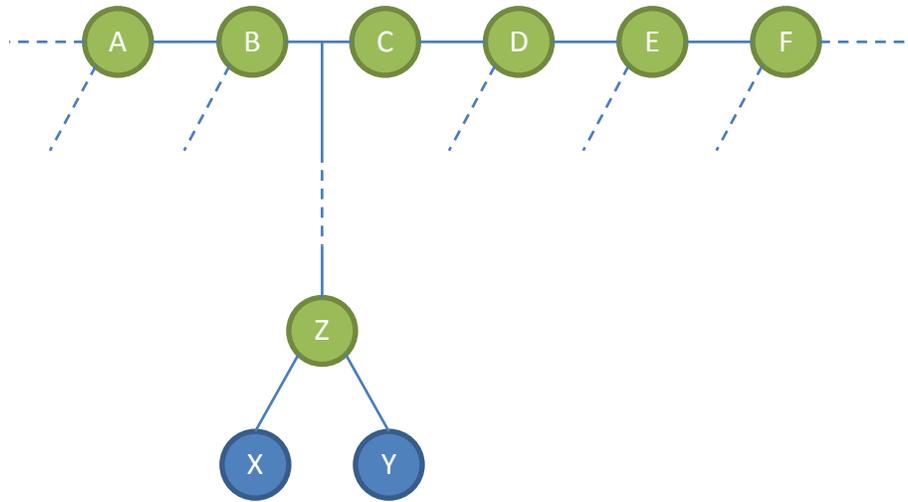

**Figure 3. The clan Z is only distantly related to BC.**

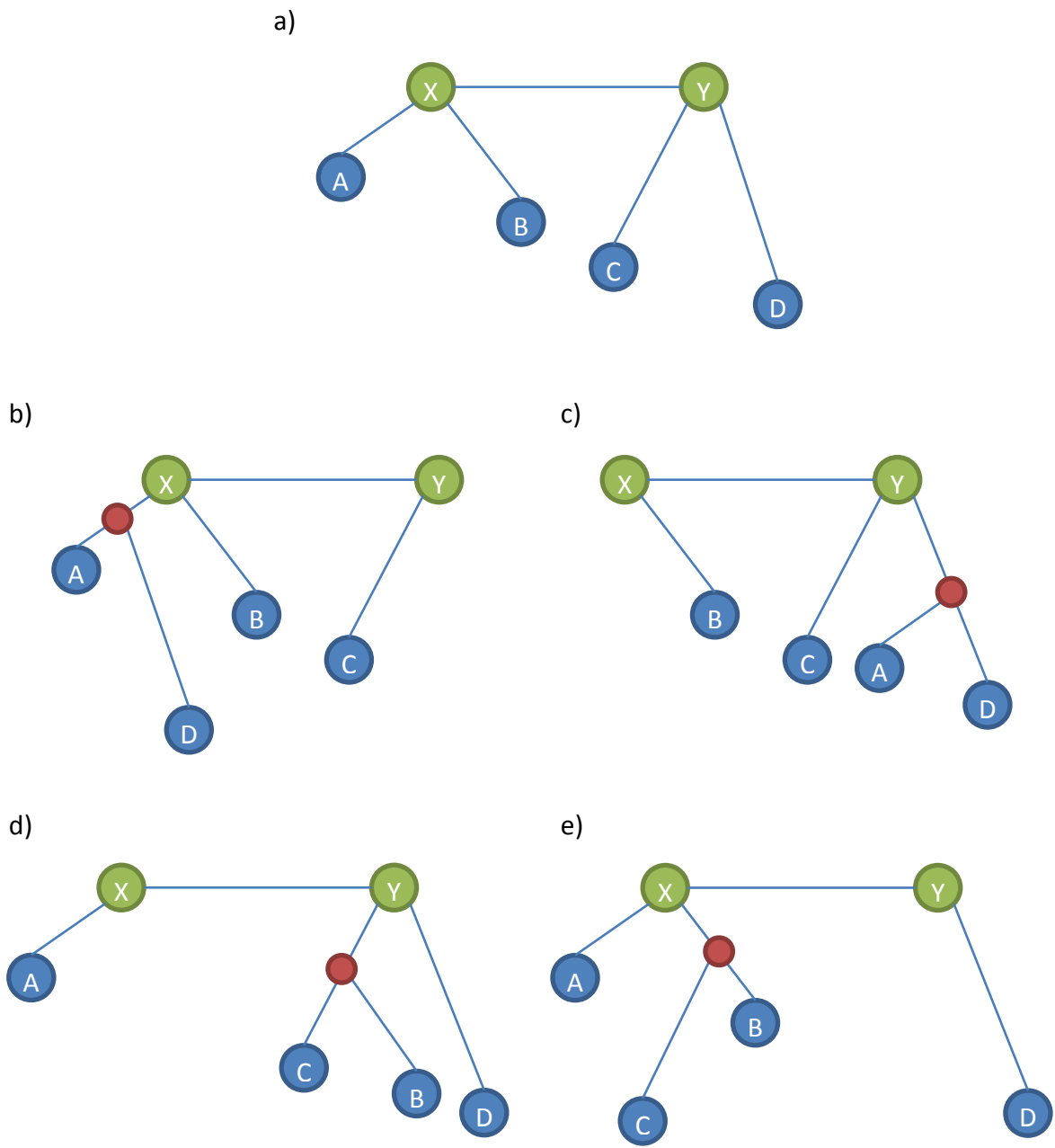

**Figure 4. a)** The tree "((A,B),(C,D));"; b-e) its four Type (ii) SPRs, the latter of which are equivalent to each other, if and only if branch lengths are not considered.

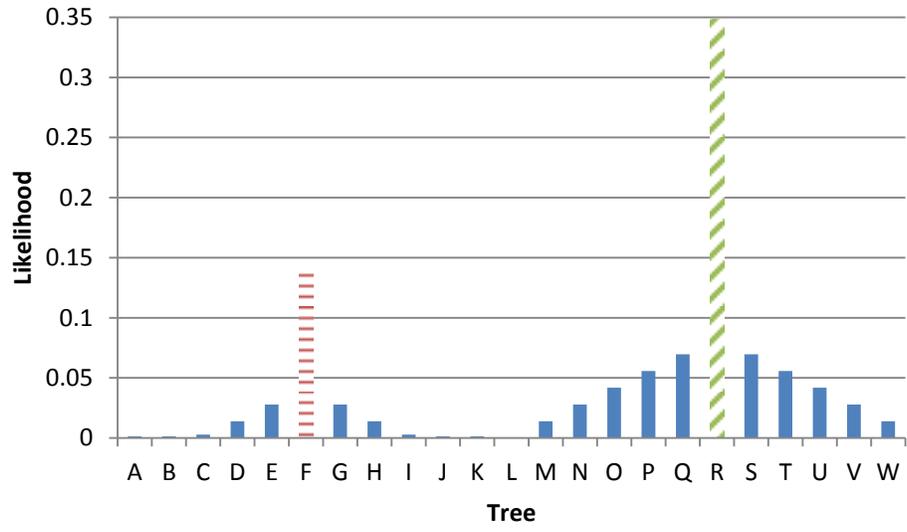

**Figure 5. Hypothetical likelihood of 25 trees.**

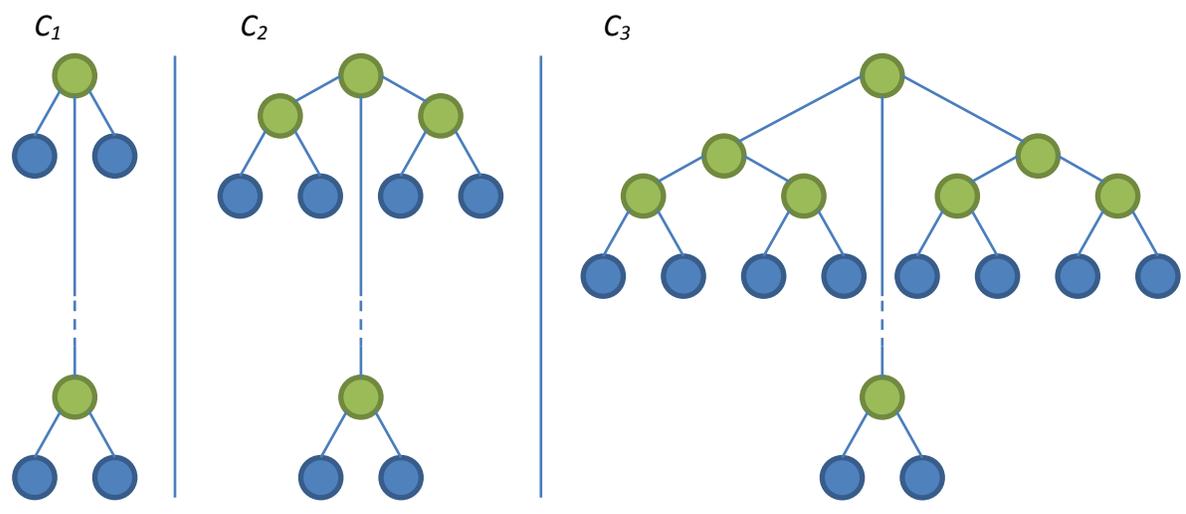

**Figure 6.** The topologies $C_1$, $C_2$, and $C_3$.

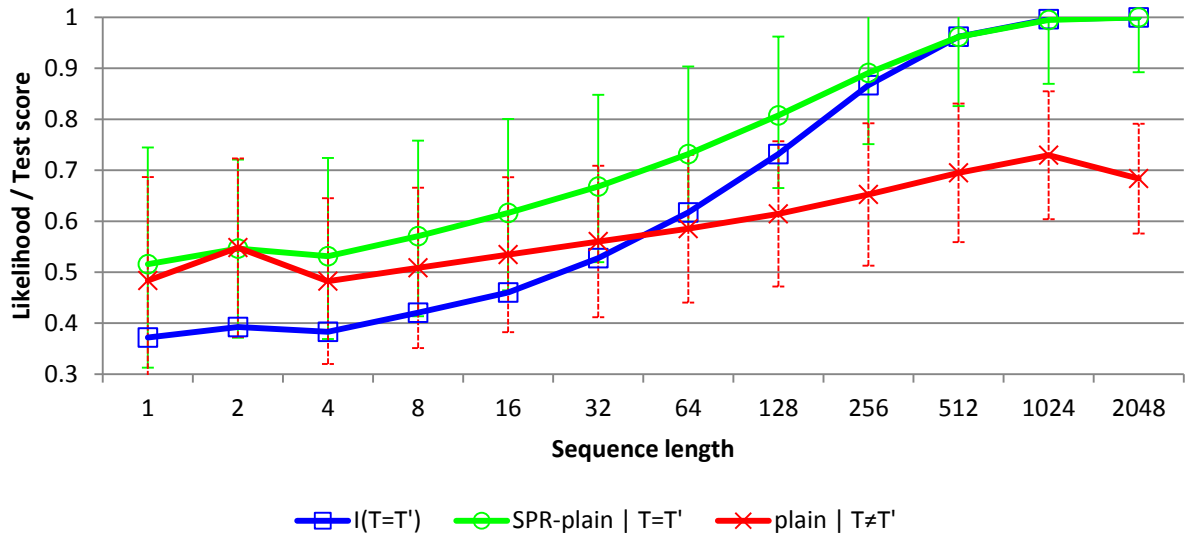

Figure 7. The likelihood of recovering the correct tree (4-species tree with unit length branches), and SPR$_{plain}$ test scores, for various sequence lengths. "SPR$_{plain}$ | T=T '" means the average SPR$_{plain}$ statistic, conditional on having identified the correct topology. Error bars show one standard deviation.

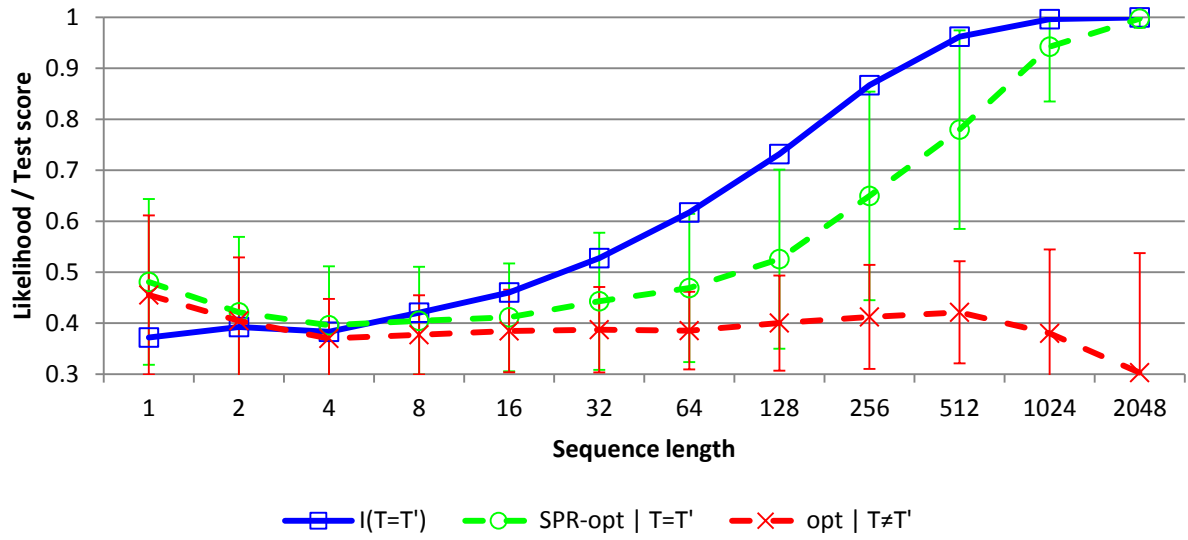

**Figure 8.** The likelihood of recovering the correct tree (4-species tree with unit length branches), and SPR$_{opt}$ test scores, for various sequence lengths.

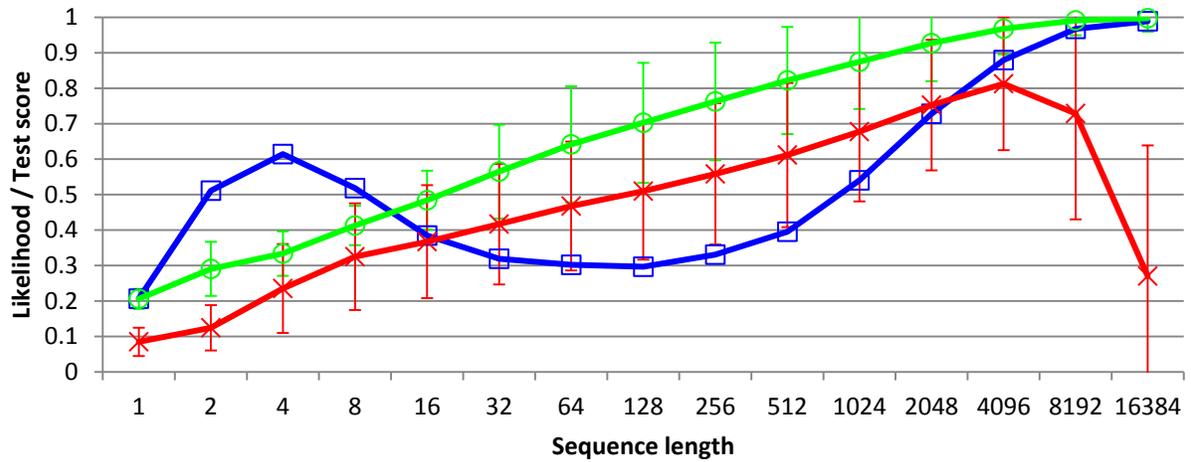
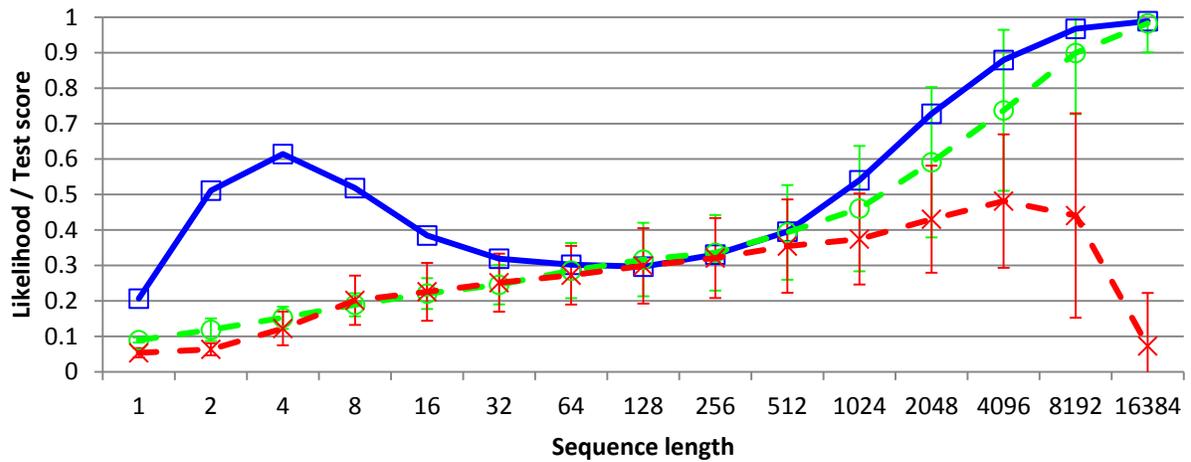

Figure 9. The likelihood of recovering the correct tree ($C_2$), and SPR test scores, for various sequence lengths.

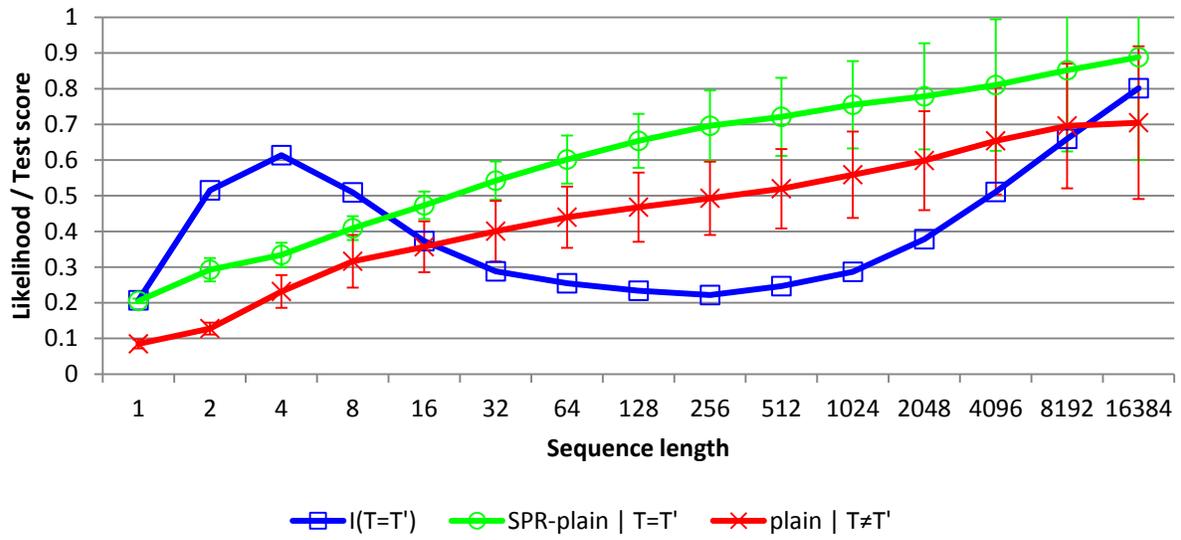
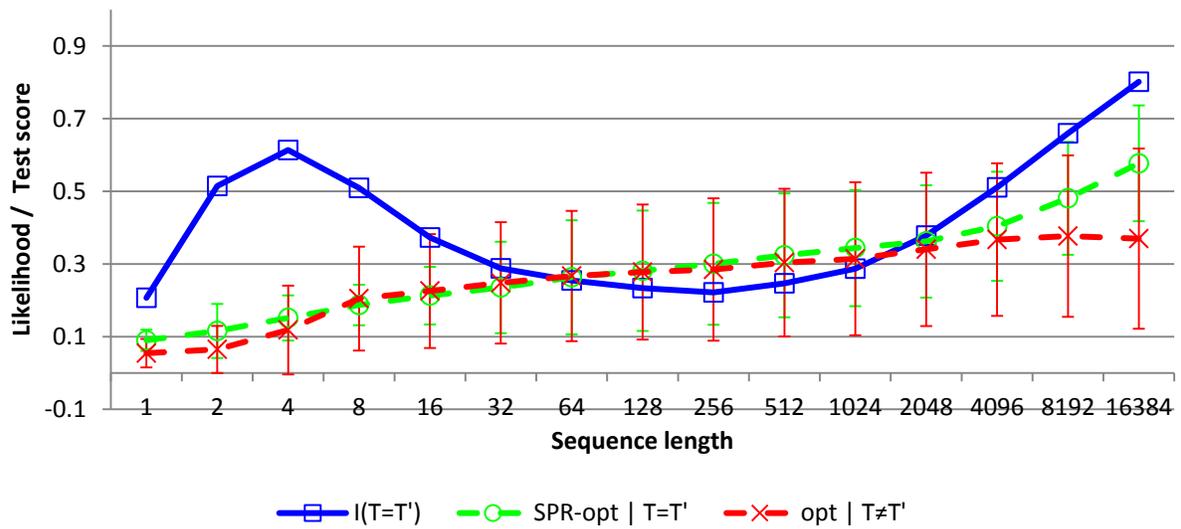

**Figure 10.** The likelihood of recovering the correct tree ($C_2$ with a very long branch), and SPR test scores, for various sequence lengths.

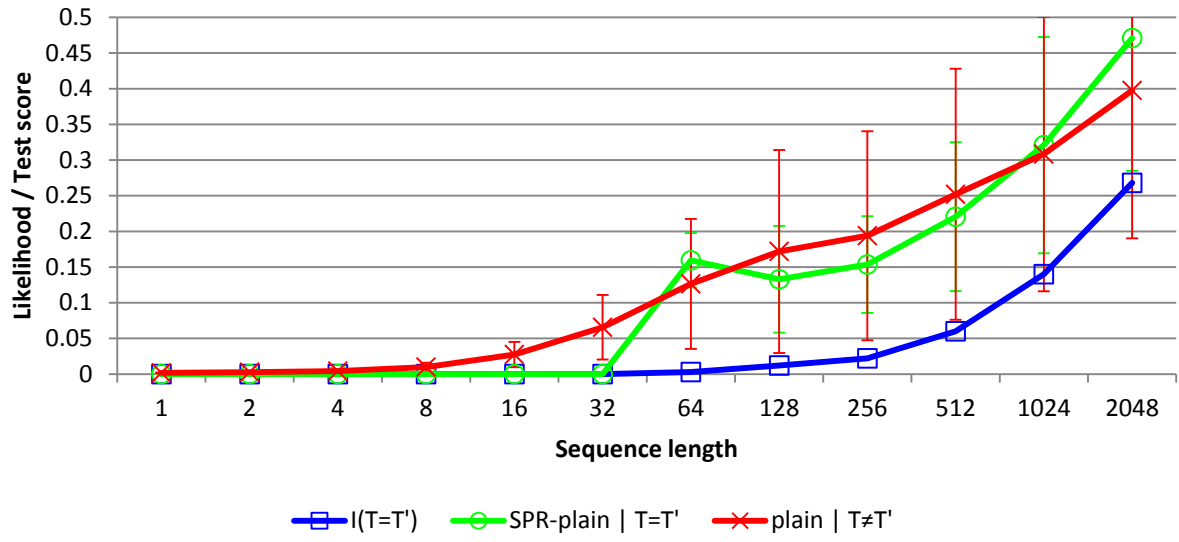

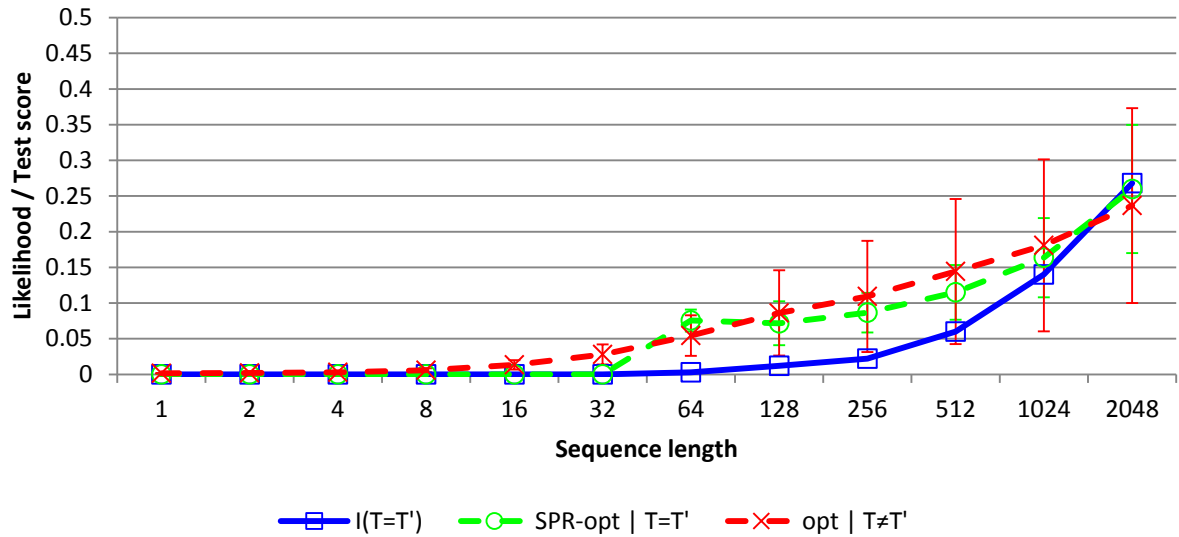

Figure 11. The likelihood of recovering the correct tree ($C_4$), and SPR test scores, for various sequence lengths.

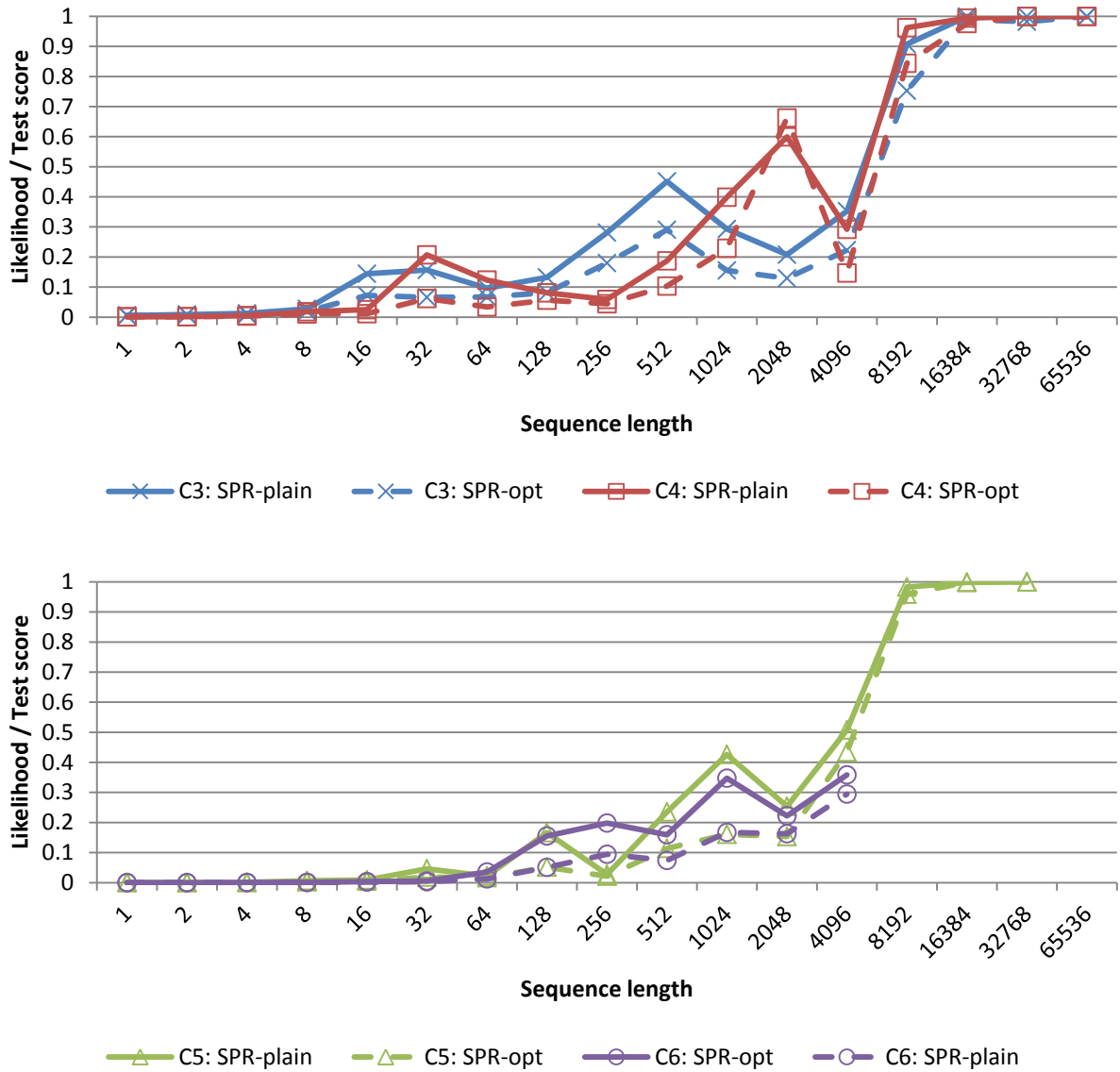

Figure 12. The SPR test scores, for various sequence lengths, with trees $C_3$, $C_4$, $C_5$, $C_6$.

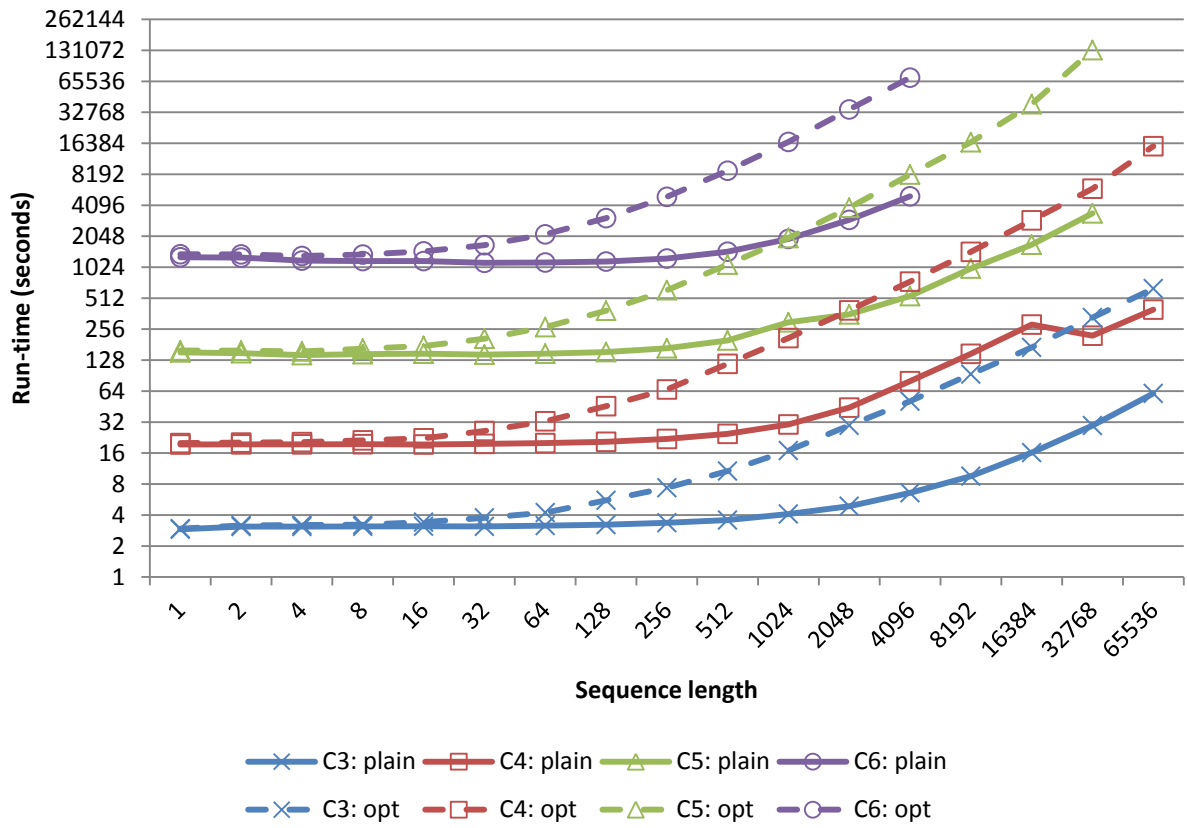

Figure 13. The run-time of the SPR tests, for various sequence lengths, with trees $C_3, C_4, C_5, C_6$.

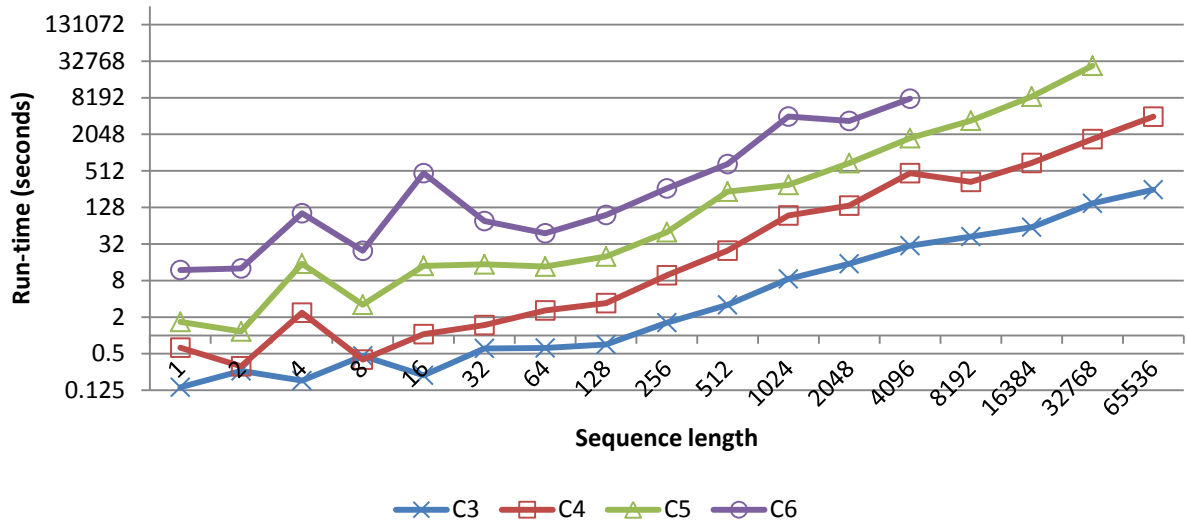

Figure 14. The run-time of DNAML, for various sequence lengths, with trees $C_3$, $C_4$, $C_5$, $C_6$.

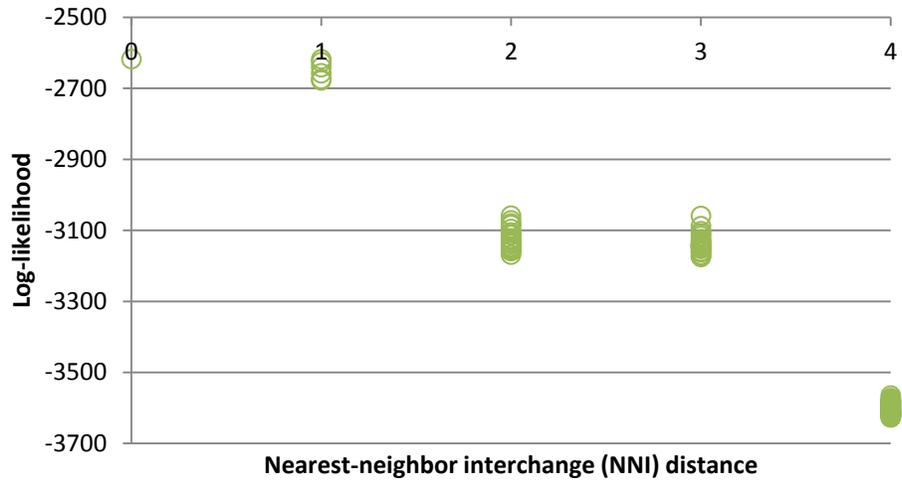

Figure 15. Log-likelihoods of the 6-species topologies, given data which are always generated from tree $C_2$.

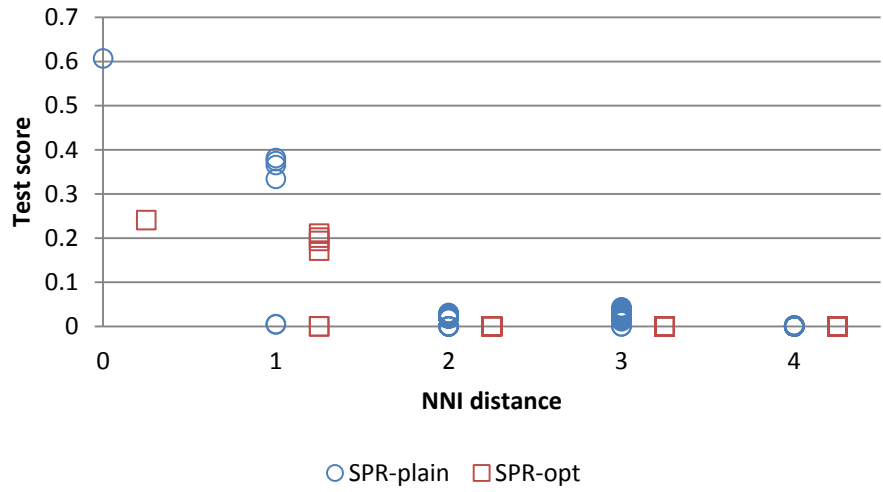

**Figure 16.** SPR statistics of the 6-species topologies, given data which are always generated from tree $C_2$. For clarity, the SPR-opt data points are shifted horizontally.